\documentclass[aps,prl,twocolumn,groupedaddress]{revtex4}
\usepackage{graphicx}
\usepackage{dcolumn}
\usepackage{bm}

\def\x{{\bf x}}
\def\y{{\bf y}}

\def\lsim{\mathrel{\rlap{\lower4pt\hbox{\hskip1pt$\sim$}}
    \raise1pt\hbox{$<$}}}
\def\gsim{\mathrel{\rlap{\lower4pt\hbox{\hskip1pt$\sim$}}
    \raise1pt\hbox{$>$}}}
\begin{document}

\preprint{IFT/12/02}

\title{ Description of the $D^*_s(2320)$ resonance as the $D\pi$ atom }

\author{ Adam P. Szczepaniak }
\affiliation{ Physics Department and Nuclear Theory Center \\
Indiana University, Bloomington, Indiana 47405 }

\date{\today}

\begin{abstract}
We discuss the possibility that the recently reported  resonance in the
$D_s \pi^0$ spectrum can be described in terms of residual $D\pi$
interactions. 
\end{abstract}
 

\maketitle

The BaBar collaboration has recently reported a narrow resonance in the
$D^+_s(1968)\pi^0$ spectrum~\cite{Aubert:2003fg}. 
 The mass of the resonance $M_r =
 2.32\mbox{ GeV}$ is significantly below the $DK$ threshold, and the 
 width $\Gamma \sim 9\mbox{ MeV}$ is of the order of a typical
 hadronic decay width for a light meson emission from a charmed resonance. 

In the charmed sector there are three, stable under hadronic decays,  
  light-flavored,  $c{\bar q}$, $q=u,d,s$,   $D$-mesons, the 
   $D^0(1870)$, $D^+(1870)$ and $D_s(1968)$ together with  
 their spin excitations with $J^P=1^-$, 
 the $D^{*}(2010)$ and the $D^{*}_s(2110)$ in the $u,d$ and strange sector 
  respectively~\cite{PDG}. 
 Other well established resonances have $J^P=1^+$, the 
 $D_1(2420)$ and the $D_{s1}(2536)$. 
 In terms of the quark model classification the ground 
 states with $J^P=0^-$ correspond to  $^{2S+1}L_J= \;^1S_0$ $c\bar q$ states,
 the $J^P=1^-$ natural parity sates are identified as $^3S_1$ states and the
 $J^P=1^+$ unnatural party resonances  are the $J=1$ members of the
 $L_{c{\bar q}}=1$  multiplet containing states with the following
 quantum numbers,  $^3P_0$, $^3P_1$, $^0P_1$ and $^3P_2$.  
 The predicted $^1P_2$ states  could be assigned to $D^{*}_2(2460)$
 and $D_{sJ}(2573)$ resonances  and  two more states, the $^3P_0$ 
 and a linear  combination of the $^3P_1$ and $^1P_1$ are still  to be found. 

As pointed out by Barnes {\it et al.} the identification of the BaBar
state with the $^3P_0$ member of the $L_{c{\bar q}}=1$ multiplet is
unlikely~\cite{Barnes:2003dj}. Its mass is $230\mbox{ MeV}$ below the average 
 of the $D_{s1}$ ($^3P_1$) and $D_{sJ}$ masses. 
 Furthermore from the heavy quark
 symmetry it is expected that two out of the four $L_{c\bar q}=1$ states,
 corresponding to the $j_{\bar q} = L_{c\bar q} +{1/2}_{\bar q} = 3/2$ doublet 
  are narrower then the other two from the $j_{\bar q} =1/2$ doublet. 
 The former can be identified with the narrow $D_{s1}$ and $D_{sJ}$ states,
  while the latter would include the $^3P_0$ state, which in a quark
  model is predicted to have width of the order of hundreds of $\mbox{
    MeV}$~\cite{Godfrey:wj}. 

To summarize, the measured charmed mesons resonances, with the
exception of the latest BaBar state seem to agree well with the quark
mode. From the point of view of this classification, 
 two states, one with $J^P=0^+$ and one with $J^P=1^+$ are missing; 
 however, they may well be much broader then those observed. 

Since the $J^P=0^+$ BaBar state is not expected to belong to
a $c\bar q$ family we investigate the possibility it is  
molecular in nature. This could happen if there is a strong
flavor-singlet attraction between the pion and the $c\bar s$
mesons. Since $m_\pi/m_{c\bar q} < 10\%$ one could consider the
BaBar state as a result of a pion being captured by a non-relativistic
(even static) charmed meson. Since the width of the resonance  
 measured by BaBar, ($\Gamma \lsim 10\mbox{ MeV}$)  is small compared to the 
 energy difference between  nearby coupled channels, {\it e.g.} 
 $|m_{D^*_s(2320)} - m^{tr}_{DK}| = 40 \mbox{ MeV}$, channels other
 than the  measured $D_s\pi$ should be unimportant. 
  
 Even though it is expected that there are residual flavor-neutral
 interactions mediated by glueball (pomeron) exchanges, the details
 of such processes are presently unknown. It is 
 possible, however to formulate the problem using effective interactions
 once the relevant energy-momentum scales have been identified. In 
 particular, the $D_s\pi$ interactions are mediated via
 multi-gluon exchange and its spectral properties at low mass can be
 saturated by $\eta'$ exchanges thus correlated with matter
 fields~\cite{Rosenzweig:1979ay}. 
 The virtual light quark matter fields coupled 
 to $\pi$ or $D$ mesons probe the light quark distribution in these
 particles up to momentum scales of the order of the QCD scale 
  $\Lambda \sim 0.5 - 1\mbox{ GeV}$, thus momenta in virtual meson
 propagators should be truncated at $p\lsim \Lambda$. The effective $D\pi$
 interaction obtained this way could then be used to calculate the
 $D\pi$ scattering amplitude~\cite{Oller:1998zr, Oller:1998hw}. 
 This requires iteration of the real part
 of virtual $D\pi$ exchanges. Since we are not explicitly including
 contributions from other channels the energies of the intermediate
 states have to be truncated at $E(p) \le E_{th}$. For 
 example for the $DK$ threshold, $E_{th} = 2.36\mbox{ GeV}$, 
  which implies the relative  momentum in the $D\pi$ system 
 $p \le 340 \mbox{ MeV}$. 
 Thus the cutoff, $\mu$ on the loop integrals
 over $D\pi$ states should be of the order of a few hundred MeV. 
 Of course if all coupled channels were explicitly included, it would
 be  possible able to set $\mu \to \infty$. 

  To summarize, the effective $D\pi$ flavor-singlet interaction should
 have a natural strength if the scale in the interaction is of the
 order of the QCD scale ($\Lambda$) and 
 the $D\pi$ amplitude is truncated at momenta
 of the order of a few hundred MeV ($\mu$). 

The effective interaction can be deduced from an effective
 QCD Lagrangian which includes anomalous $U_A(1)$ symmetry 
 breaking~\cite{Rosenzweig:1979ay}. 
 For the system under study, the relevant part of such a Lagrangian is 
 given by~\cite{DiVecchia:1980ve, Bass:1999is, Bass:2001zs} 
\begin{widetext}
\begin{equation}
{\cal L} = {{f_\pi^2} \over 4} Tr\left( \partial^\mu U \partial_\mu
  U\right) 
 + {{f_\pi^2} \over 4} Tr M\left( U + U^{\dag} \right) + {1\over 2} i Q Tr
  \left[ \log U - \log U^{\dag}\right] + {3\over { m_\eta^2 f_\pi^2}} Q^2 
 + Q^2 
 \left[  {{9\beta}\over {2f_\pi^2 m_\eta^4}} 
  Tr\left( \partial^\mu U \partial_\mu
  U\right) +   {c \over {f_\pi^4}} D^2  \right].  \nonumber \\
\end{equation}
\end{widetext} 
 The first two terms represent the lowest order terms of a nonlinear 
 chiral Lagrangian, with $U = \exp(i\pi^a T^a/f_\pi + i\sqrt{2/3}
 \eta_0/f_\pi)$, $\pi^a$ and $\eta_0$ being the octet and singlet
 meson fields, respectively. We have neglected small terms which
 differentiate  between the flavor octet, $f_\pi$ and the flavor 
 singlet meson decay  constants.  The $Q=(\alpha/4\pi) F \tilde{F}$ 
 represents the gluon field and the
 term linear in $Q$ is responsible for the anomalous coupling of the gluon to
 matter fields and for the $U_A(1)$ symmetry breaking. The
 first $Q^2$-dependent term can be interpreted as the kinetic term of 
 the gluon field. Finally the last two terms represent flavor-singlet, 
 lowest dimension gluon coupling to the light meson octet and the
 charmed, $D$ meson field.  The coupling constant $\beta=-0.63$ can be 
 determined from the
 $\eta'\to \pi\pi\eta$ decay~\cite{DiVecchia:1980ve, Bass:2001zs} 
 and, as discussed above,  the unknown coupling $c$, is expected to be 
 of the order of $\Lambda^{-2}$.  Using the equations of motion,  the 
 $Q$-field can be replaced by the matter $\eta_0$ field which 
 among others leads to the following interactions, 
\begin{equation}
{\cal L}_{\pi\pi\eta_0\eta_0} = {3\over 2} {\beta\over {f_\pi^2}} 
 \eta_0^2 \partial_\mu \pi^a \partial^\mu \pi^a,\;\;
{\cal L}_{DD\eta_0\eta_0} = {c\over 6} {{m_\eta^4}\over {f_\pi^2}}
\eta^2_0 D^2.  
\end{equation}
 These result in an effective $D\pi$ Lagrangian  given by 
\begin{eqnarray}
& & L_{D\pi} = c{\beta \over 4} {{m_\eta^4}\over {f_\pi^4}}
 \int dxdy (\partial_\mu \pi^a(x))^2 \langle T \eta_0^2(x)
 \eta_0^2(y) \rangle D^2(y).  \nonumber \\
\end{eqnarray}
The expectation value of the $\eta$ (gluon) field 
 is replaced by an instantaneous contact term, smeared over the QCD
 scale,$\Lambda$,  
 $\langle T\eta^2_0(x) \eta^2_0(y) \rangle \sim -\delta(x^0-y^0) 
 \Lambda^4 \delta^3_\Lambda (\x - \y)/m_\eta^4$, resulting in a final $D\pi$
 effective potential,  
\begin{eqnarray}
& & V_{D\pi} =  {{\beta c}\over {4f_\pi^4}} 
 \int d\x (\partial_\mu \pi^a(\x))^2  D^2(\x),  
 \nonumber \\ \label{v}
\end{eqnarray}
with $c = O(\Lambda^2) \sim O(1\mbox{ GeV}^2)$. Due to absence of
 multi-particle, relativistic effects and the low momentum approximation
($p\lsim \mu$),  the  scattering amplitude can be 
  determined from $(1-VG)^{-1}V$ with  
 $G=(E - \sqrt{m_D^2 + p^2} - \sqrt{m_\pi^2 + p^2} 
 + i\epsilon)^{-1}$ being the free $D\pi$ propagator. The scattering
 $S$-wave  phase shift can then be easily calculated for the potential
 of  Eq.~(\ref{v}) and is given by, 
\begin{equation}
 \tan\delta(E)  = -
 {{E^2_\pi(p) p  c\beta f(p/\mu) } \over {32\pi f_\pi^4 E(p) (1 -
 J[E(p)])}}, \label{ph}
\end{equation}
 with $E(p) = E_\pi(p) + E_D(p) 
 = \sqrt{m_\pi^2 + p^2} + \sqrt{m_D^2 + p^2}$ and
$J(E)$ being  the contribution from the real part of $D\pi$ loop 
 cutoff by a form factor $f(p/\mu)$ with $\mu = O(\mbox{ few 100
 MeV })$,  
\begin{equation}
J(E) = {{c\beta} \over {32\pi^2 f_\pi^4}}  \int_0^\infty dk
  {{k^2 E^2(k) f(k/\mu)}\over
  {E_\pi(k)E_D(k)\left[ E(p) - E(k) + i\epsilon\right]}} \label{J}. 
\end{equation}
 The comparison between our theoretical prediction and the BaBar result
 is shown in Fig.~1. Instead of plotting the data,  for simplicity we 
 plot a phase shift resulting from a Breit-Wigner (BW) parameterization
 of a resonance with mass $M_r = 2.32\mbox{ GeV}$ and width, 
 $\Gamma_r = 10\mbox{ MeV}$ (equal to the experimental resolution of the BaBar
 measurement).  We recall that a resonance phase shift, 
 parametrized by a simple (without energy dependence in the width)
 Breit-Wigner  formula gives, 
\begin{equation}
\sin^2\delta_{BW}(E) = {{ (\Gamma_r M_r)^2}\over {{\left[ (E^2 - M^2_r)^2 +
        (\Gamma_r M_r)^2\right] }}}, 
\end{equation}
where $M_r$ and $\Gamma_r$ are the mass and width of the resonance,
respectively. In Fig.~1 this is shown by the shaded region, whose
 size was fixed to $\Delta\sin^2\delta_{BW} = 0.1$ roughly
 corresponding to size of the errorbars in the mass distribution 
of the $D_s\pi$ events shown in Ref.~\cite{Aubert:2003fg}.  
 The prediction for $D\pi$ phase shift from Eq.~(\ref{ph}) is shown 
 with the solid line and it was calculated  using $c=1\mbox{ GeV}^2$
 and  $\mu = 341\mbox{ MeV}$. 

 \begin{figure}
 \includegraphics[width=2.5in]{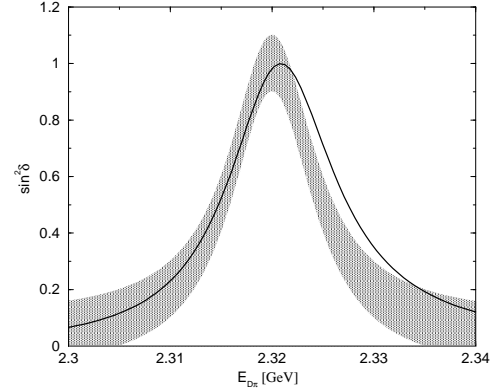}
 \caption{\label{fig1} 
 Comparison between the phase shift calculated from the formula in
 Eq.~(\ref{ph}) (solid line) with the Breit-Wigner resonance 
  with $M_r = 2.32\mbox{ GeV}$ and $\Gamma_r = 10\mbox{ MeV}$. 
 The form factor in Eqs.~(\ref{ph}) and ~(\ref{J}) was chosen as $f(p/\mu) =
 1/(1+(p/\mu)^2)^2$. }
 \end{figure}

 Since the resonance is narrow it
 is clear that the position and width will be sensitive to these
 parameters. For example with $c$ fixed  changing $\mu$ by $\pm 20\%$ 
 shifts the position of the resonance between $2.257$ and $2.393 \mbox{
   GeV}$ and $\Gamma$ decreases for low $M_r$ to $7\mbox{
   MeV}$, as the resonance mass approaches  the $D\pi$ threshold,  and 
 increases to $22\mbox{ MeV}$ at the high mass. 
 However, by changing both $c$ and $\mu$ within their natural
 ranges it is possible to restore the original resonance parameters. 
 The increasing discrepancy between the BW parameterization and the
 solid line at higher mass is due to absence of phase space factors
 (demanded by unitarity) in the BW parameterization. 

In summary we have found that using reasonable assumptions regarding
 flavor-independent interactions between the pion and the
 charmed-strange mesons, with natural parameters it is possible to 
 reproduce a narrow resonance in the $D\pi $ spectrum. 
 Such states should also be present in other charge modes, {\it
 e.g.}  $D_s\pi^\pm$.  We have also checked that our findings are 
 insensitive to the   details of a formulation, {\it e.g.} 
 we studied the nonrelativistic
 approximation and used the $N/D$ method~\cite{Oller:1998zr}.

\begin{table}
\caption{\label{tab:table1}
 Predictions for the $J^{P}=0^+$ $c{\bar u}({\bar d})$ ($D_0$) and 
 charmed-strange $c{\bar s}$ ($D_{s0}$) meson masses and widths obtained
 with $c=1$ and $\mu = 340 \pm 68\mbox{ MeV}$. }
\begin{ruledtabular}
\begin{tabular}{lcr}
 & $M_r$ [GeV]  & $\Gamma_r$ [MeV] \\ \hline
$D_0$    & 2.15 - 2.30 & 7-24\\
$D_{s0}$ & 2.44 - 2.55 & 17-42\\
\end{tabular}
\end{ruledtabular}
\end{table}

  A similar analysis applied to the $J^P=0^+$ $D\pi$ and $DK$
 systems produces scalar resonances with masses and widths listed in
 Table.~1. These predictions should be easily tested by experiment
 because of the narrow width of the states involved. 
 These masses are comparable with the quark model
  predictions of $M_{D_0} = 2.4 \mbox{ GeV}$ and $M_{D_{s0}} =  2.48
 \mbox{ GeV}$ respectively~\cite{Godfrey:xj}, however none of these 
 states have been observed yet. We have also found that the observed 
  $D^{*}_1(2420)$, and $D_{s1}(2536)$ can be generated in the $D^*\pi$ 
 and the $D^*K$ systems using a similar mechanism. Resonances in
 $D^*\pi$ or  $D^*K$ could in principle 
 be studied this way as well; however, since the lifetimes of the $D^*$'s are 
 comparable to that of the expected two-meson resonance 
 the breakup channels of the $D^*$ would have to be included
 explicitly and those may prevent from narrow resonance in the
 two-meson channels to be formed in the first place. 
 This is also true for possible molecular states build around ${c\bar c}$
  mesons which can annihilate through strong interactions. 
 Finally the interaction of Eq.~(\ref{v}) also leads to
 interactions in the relative $P$-wave of the two-meson system, however
 the resulting phase shift is slowly varying and does not 
 display resonance characteristics. 
 
 It is a pleasure the acknowledge Alex Dzierba, Don Lichtenberg, Dan Krop and
 Scott Teige for several discussions. This work was supported in part
 by US DOE contract, DE-FG02-87ER40365.

\end{document}